\documentclass{IEEEtran} 
\usepackage{graphicx}
\usepackage{hyperref}
\usepackage{amsmath,amssymb}
\usepackage{subfig}
\usepackage{algorithm}
\usepackage{algorithmic}
\usepackage{multirow}
\usepackage{array}
\usepackage{balance}
\usepackage{url}
\newtheorem{definition}{Definition}
\DeclareCaptionType{copyrightbox}
\newcommand{\qed}{\hfill \ensuremath{\Box}}

\begin{document}

\title{``You Know Because I Know'': a Multidimensional Network Approach to Human Resources Problem}

\author{
\IEEEauthorblockN{Michele Coscia$^{1}$, Giulio Rossetti$^{2}$, Diego Pennacchioli$^{2,3}$, Damiano Ceccarelli$^{4}$, Fosca Giannotti$^{2}$\\}
\IEEEauthorblockA{
$^{1}$ CID - Harvard Kennedy School, 79 JFK Street, Cambridge, MA, US, Email: michele\_coscia@hks.harvard.edu\\
$^{2}$ KDDLab ISTI-CNR, Via G. Moruzzi 1, Pisa, Italy, Email: \{name.surname\}@isti.cnr.it\\
$^{3}$ IMT Institute for Advanced Studies, Piazza San Ponziano 6, Lucca, Italy\\
$^{4}$ Navionics, Lucca, Italy, Email: dceccarelli@navionics.com\\
}
}

\maketitle

\begin{abstract}
Finding talents, often among the people already hired, is an endemic challenge for organizations. The social networking revolution, with online tools like Linkedin, made possible to make explicit and accessible what we perceived, but not used, for thousands of years: the exact position and ranking of a person in a network of professional and personal connections. To search and mine where and how an employee is positioned on a global skill network will enable organizations to find unpredictable sources of knowledge, innovation and know-how. This data richness and hidden knowledge demands for a multidimensional and multiskill approach to the network ranking problem. Multidimensional networks are networks with multiple kinds of relations. To the best of our knowledge, no network-based ranking algorithm is able to handle multidimensional networks and multiple rankings over multiple attributes at the same time. In this paper we propose such an algorithm, whose aim is to address the node multi-ranking problem in multidimensional networks. We test our algorithm over several real world networks, extracted from DBLP and the Enron email corpus, and we show its usefulness in providing less trivial and more flexible rankings than the current state of the art algorithms.
\end{abstract}

\section{Introduction}
Finding talents is one among the most difficult challenges for organizations. Hiring talents means performing better, get more revenue and evolve the business. Where hiring talents is relatively easy, the biggest challenge for organizations today is to find talents they have already hired: finding and creating knowledge is important, but so it is to be able to search and mine knowledge that is already owned. The social networking revolution allowed the creation of tools to evaluate competencies, expertise and skills, like Linkedin\footnote{\url{http://www.linkedin.com/}}. Before social networks, talent management activities were restricted to a marketing oriented approach: to discover talents among their employees, organizations promote internal contests or invest into assessment activities and campaigns. However, during the last few years, organizations started also to put efforts on social talent management, often connected to wider social related initiatives (social CRM, enterprise social networking, etc.). Social talent management is nowadays based on dedicated pages or applications whose aim is to discover interesting professionals. Social networks are also used by organizations to get unofficial information about their employees or candidates, to understand what they do and who they really are.

To understand where and how an employee is positioned on a skill network will enable organizations to find previously hidden sources of knowledge, innovation and know-how. Resumes can provide structured information about studies and working experiences, but they are not useful to understand skills and experiences that do not belong directly to the employee, but to his friends and colleagues. If a candidate does not master a topic but has a strong relationship with somebody who does, then he is an important gateway in that topic, although different relationships allow for different gateway values (i.e. two friends in the same company represents a stronger connection than two friends in competing companies).

If an employee is involved on specific tasks, and she has always been involved only on those tasks, this does not mean she could not have strong competencies and skills on completely different subjects: hobbies, passions, interests can be worth some, sometimes a lot of, value in the \textit{prosumers} age. People are digitally involved throughout their life and there is not a clear separation between personal and professional network. Understanding the position of somebody among different skills and knowledge networks or rankings can let this value emerge. This data richness and hidden knowledge demands for a multidimensional and multiskill approach to the employee ranking problem: the definition of a ranking algorithm on networks, able to capture the role of different kinds of relations and the importance of different skill sets.


Given a person with a set of skills and a neighborhood of friends in a social network, the skills of the friends to some extent are accessible through that person, and therefore they should be considered when evaluating her. More formally, each node $n$ in a network has some skills $S$ each with a given intensity, and it is connected, with different kinds of edges, to other nodes ($n_2, n_3, ...$) with their own skill set ($S_2, S_3, ...$). The real value of node $n$ is then defined by a function $f$ that takes as input not only $S$, but also $S_2, S_3, ...$, by accounting for the different dimensions connecting $n$ to $n_2, n_3, ...$. This idea has been proven in the economic field at the macro level: in \cite{moretti} the author proved that the social return of higher skill levels is higher than the personal return, i.e. higher skilled people make their colleagues to be more valuable as well.

Current ranking algorithms can only provide multiple rankings on monodimensional networks, or simple rankings in multidimensional networks. In this paper we propose an algorithm whose aim is to provide multiple rankings (one for each skill) for nodes in a multidimensional network, a network with multiple types of edges. Our approach is called ``you (U) know Because I Know'', or UBIK. We test UBIK on real world networks, showing that its ranking is less trivial and more flexible than the current state-of-the-art methods.

Our contribution can be then summarized as follows. We introduce a novel ranking algorithm for multidimensional networks with multiple node attributes, able to provide a different ranking for each skill. We provide a fast implementation able to scale linearly in the number of edges of the network. We provide a new ranking for real world scenarios, including co-authorship in computer science and corporate emails.

In Section \ref{sec:related} we present the related works. The intuition and the formal model behind this work is explained in Section \ref{sec:idea}, while the actual implementation of the UBIK algorithm is described in Section \ref{sec:UBIK}. We provide our experimental setting in Section \ref{sec:experiments}. Finally, Section \ref{sec:conclusion} concludes the paper.

\section{Related Works}\label{sec:related}
How to evaluate and manage the human resources of an organization is an endemic problem in management. We are not the first researchers pointing out that social network analysis is a useful tool in this area, as \cite{hr-sna} starts from the assumption that the value of a person is not only determined by the extent of what she knows, but also to her position in a social network, validating our starting point. However, \cite{hr-sna} only uses a centrality concept, without looking at the skills of the nodes of the network, nor it builds a computer science framework to solve the problem. The social dimension enters in the problem of human resources management not only as an evaluation tool of the skills of a person, but also on how much she can influence the behaviors of other employees \cite{hr-influence}.

To rank nodes according to their importance in a network structure is a classical problem in complex network analysis \cite{wagner}. Our final aim is to have multiple rankings over a multidimensional network. To the best of our knowledge, no algorithm is able to perform this task. The current state of the art in this research branch can be divided in four categories.

In the first category, there are popular ranking algorithms (PageRank, SALSA and related \cite{pagerank,salsa,sidiropoulos}) which provide a single ranking in a monodimensional network.

In the second category, we have ranking algorithms that can provide multiple rankings, but on a non-multidimensional network. The oldest and best-known approach is HITS \cite{hits}, which provides only two rankings (hub and authorities). Another example is topic-sensitive PageRank \cite{ppr}, that can provide an arbitrary number of rankings.  However, in the topic-sensitive PageRank there is no way to exploit different kinds of relations, so it cannot perform the task requested.

In the third category, the ranking algorithms deal with multidimensional network, but they provide a simple ranking, much like the PageRank does for monodimensional networks. This branch has been thoroughly tackled in recent years, for example in \cite{multirank, har, topic-level}. Since we are interested in a multi-ranking, none of these works can help us.

Finally, this work aims to be included in the fourth category, multidimensional approaches to the multi-ranking problem. There is already one method in literature in this category, namely TOPHITS \cite{tophits}. However, just like HITS, TOPHITS provides only two different rankings, hubs and authorities, therefore its application to real world scenarios in the evaluation of many skills is questionable. With this aim in mind, we proceed explaining our methodology.

\section{Network-Based Human Resources}\label{sec:idea}
\subsection{Problem Definition}\label{sec:blocks}
As stated in Introduction and Section \ref{sec:related}, our paper aims to tackle the problem of multiple rankings in multidimensional network. Our problem definition is the following:

\begin{definition}\label{def:problem}
Let $G=(V,E,D,S)$ be a multigraph where each node $v \in V$ is connected to its neighbors through multiple edges $e \in E$, each carrying a label $d \in D$; and $S$ be a skill set, such that each node $v$ is labeled with one or more skills $s \in S$, each with a given weight $w \in R^+$. Given a query $q$ containing a set of skills $S_q \subseteq S$ and the importance $r(d) \forall d \in D$, we want to rank nodes accordingly to the weight of each $s \in S_q$ they posses directly or indirectly through their connections. \qed
\end{definition}

The intuition behind our idea is the following. Suppose we have a set of people, each with her own skills and acquaintances, and a task to be performed. In a world without social knowledge interaction, the best way to perform the task is to assign it to the person, or to a set of people (i.e. a team), possessing the highest value of the related skill. However, each person can access to the external knowledge of their acquaintances, thus possibly modifying her skill set value, and therefore the decision of the composition of the team. Each person can also access to the acquaintances' acquaintances skills, but with an increasing cost at each degree of separation, causing at some point the external skill to be useless.

Now, we need to define the social connections, and their different types. We need to formally define the skills carried by each individual as the initial state of the system and how the expertise propagates through social connections.

\subsection{The Model}\label{sec:model}
Following \cite{multidim-asonam} we model our problem with a multidimensional network (also studied in \cite{kazi-net}). We now present the basics of multidimensional networks and then the extensions we apply to the basic model.

In the case of a multidimensional setting, a convenient way to model a network is a \emph{labeled multigraph}. Intuitively, a labeled multigraph is a graph where both nodes and edges are labeled, and where there can exist two or more edges between two nodes. Just as any regular labeled graph, also labeled multigraphs may be directed and undirected, thus we allow edges to be both directed and undirected, when the analytic aim requires it. Such a graph is denoted by a triple $G=(V,E,D)$ where: $V$ is a set of nodes; $D$ is a set of labels representing our dimensions; $E$ is a set of labeled edges, i.e., it is a set of triples of the form $(u,v,d)$ where $u,v \in V$ are nodes and $d \in D$ is a label. We assume that given a pair of nodes $u, v \in V$ and a label $d \in D$ it may exist only one edge $(u,v,d)$.

In \cite{multidim-asonam}, the previous paragraph fully describes a multidimensional network. We need to add several things in order to fit into our problem definition. First, we need to introduce weighted node labels. In other words, each node $v \in V$ is a collection of couples in the form $(s, w)$ where $s$ is the label and $w \in R^+$ is the value of label $s$ for node $v$. Therefore, $v = \{(s_1, w_1), (s_2, w_2), \ldots, (s_n, w_n)\}$. The set of node labels describes what in our data are the skills of the node, along with their value. The set of all possible skills is fixed for the network, and we refer to it as $S$.

In \cite{multidim-asonam}, dimensions are considered distinct but equal. In our case, each dimension can have a different importance: in the real world a friendship tie may be more or less strong than a working collaboration, given the social environment where this tie may play its role. Therefore, each dimension $d \in D$ is represented not only by its label, but also by a value $r(d) \in R^+$, quantifying how much relevant is a relation expressed in dimension $d$, according to the query requested.

At this point we have all the building bricks of the static part of our model. Next, we define how the knowledge exchange dynamics takes place in the model itself, generating the flow that allows us to rank the nodes in the network. The idea is that each node passes the entire set of its skills to each one of its neighbors. This procedure is similar to the one employed by the classical PageRank formulation, but with a few distinctions. First, our method is not based on random walks, but on the percolation of the various skills without random jumps. Second, the amount of skill value passed to the neighborhood of a node is not equally divided and assigned to each neighbors. The amount of value that node $u$ passes to the neighbor $v$ is proportional to the importance of the dimensions connecting $u$ and $v$. It is also inversely proportional not only to the degree of the node passing the skill, but also to the degree of the node receiving the skill. Third, the knowledge exchange may be or may be not mutual, i.e. we are not narrowing our model only to directed graphs, but to general graphs.

\begin{figure}
\centering
\footnotesize
\begin{tabular}{c|l|c|}
\cline{2-3}
\multirow{13}{*}{\includegraphics[width=.25\columnwidth]{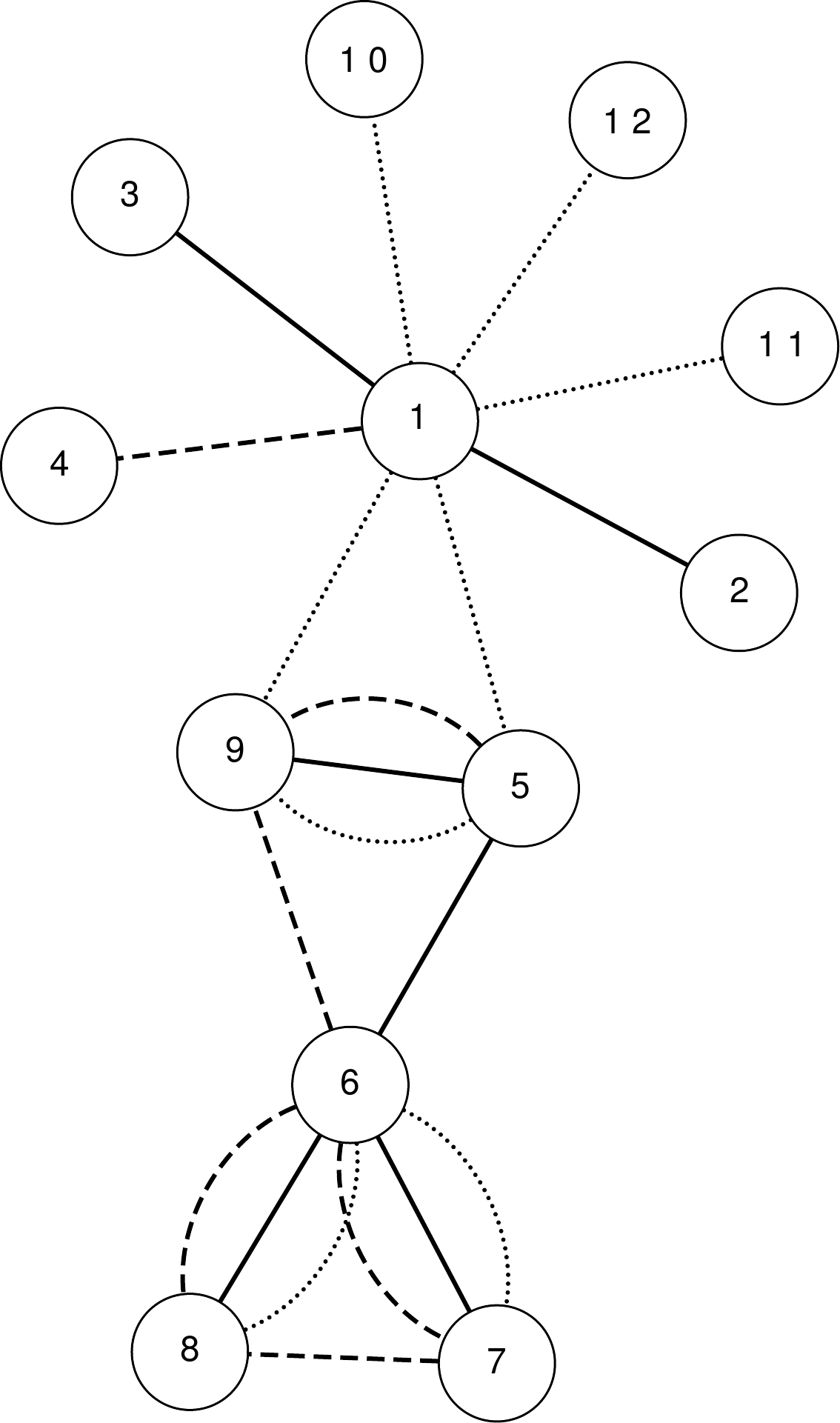}} \qquad\qquad &
Node & Skill-Value\\
\cline{2-3}
& 1 & ${(a, 115), (b, 0), (c, 0), (d, 0)}$\\
& 2 & ${(a, 100), (b, 0), (c, 0), (d, 0)}$\\
& 3 & ${(a, 100), (b, 0), (c, 0), (d, 0)}$\\
& 4 & ${(a, 100), (b, 0), (c, 0), (d, 0)}$\\
& 5 & ${(a, 100), (b, 0), (c, 0), (d, 0)}$\\
& 6 & ${(a, 0), (b, 0), (c, 0), (d, 80)}$\\
& 7 & ${(a, 0), (b, 90), (c, 20), (d, 0)}$\\
& 8 & ${(a, 0), (b, 20), (c, 90), (d, 0)}$\\
& 9 & ${(a, 100), (b, 0), (c, 0), (d, 0)}$\\
& 10 & ${(a, 100), (b, 0), (c, 0), (d, 0)}$\\
& 11 & ${(a, 100), (b, 0), (c, 0), (d, 0)}$\\
& 12 & ${(a, 100), (b, 0), (c, 0), (d, 0)}$\\
\cline{2-3}
\multicolumn{3}{c}{}\\
\multicolumn{1}{c}{(a)} & \multicolumn{2}{c}{(b)}\\
\end{tabular}
\caption{Our toy example. (a) The multidimensional network structure (each line style represents a different network dimension; (b) The skill table, recording the values of each skill ($a$, $b$, $c$ and $d$) for each node.\vspace{-0.5cm}}
\label{fig:toy-example}
\end{figure}

We use also the range parameter $\alpha$, commonly used for centrality scores, handling the following situation. If $u$ and $v$ are connected through a node $v_2$, then the amount of knowledge they exchange is lower than a direct connection, just like the resistance loss in an electric circuit. If $x$ is the amount of value passed by a direct connection, the amount of skills received from nodes $\ell$ degrees away is corrected as $x^{\frac{1}{\ell \alpha}}$. Traditionally, directly connected nodes should be at zero degrees of separation, because no other \textit{nodes} should be crossed to reach them. For practical purposes, in this paper we assume $\ell = 1$ for neighboring nodes, as we need to cross one \textit{edge} to reach them. The introduction of $\alpha$ is due both to logical and practical reasons. Logically, it makes our model more realistic, as a person is a gateway of her friends' skills, thus she is to some extent also a gateway for her friends' friends' skills, but of a much lower importance. Without the $\alpha$ parameter the skill percolation could potentially continue indefinitely, and the computation may not be able to stop.

In Figure \ref{fig:toy-example} we represented a simple toy example. The network structure of social connections is depicted in Figure \ref{fig:toy-example}(a), while Figure \ref{fig:toy-example}(b) is the skill table associated with the structure. From the skill table we know that all nodes start with some global skill value (not equal for everybody, as it happens in reality), distributed along four skills ($a$, $b$, $c$ and $d$). The social connections do not have all the same value: solid line has a $50\%$ efficiency in the knowledge transfer, dashed line has a $33\%$ efficiency, while the dotted line has only a $17\%$ efficiency. By looking only at the network structure, node 1 is the most central node. It also has the highest global skill value. According to the closeness centrality, also nodes 5 and 9 are more central than 6. However, our algorithm will propagate skills $a$, $b$ and $c$ to 6 with the maximum efficiency, while 6 will retain also its unique $d$ skill. At the end of the process, node 6 ends up as the most valuable node in general in the network, while node 1 can only specialize in skill $a$. With relaxed values for the $\alpha$ parameter (like $\alpha = 1$) node 1 can still get some parts of skill $d$ ($\sqrt{\frac{1}{6}\frac{1}{2}80}$ from node 5 and $\sqrt{\frac{1}{6}\frac{1}{3}80}$ from node 9, that is $\sim 4$.$69017$) and even less of $b$ and $c$. If $\alpha = 3$ then the contributions to node 1 of skills different from $a$ is negligible.

\subsection{The Data}\label{sec:data}
Our model is describing, according to our hypothesis, how knowledge flows in a face-to-face social environment, following the proven macro level mechanism of the social effect of schooling \cite{moretti}. However, the data about the face-to-face interactions are usually part of the tacit realm of knowledge. If we cannot find direct or proxy data sources about these interactions, any algorithm solving the problem of evaluating people on the basis of our hypothesis is practically useless. In this section, we present how we use two real-world datasets, adapting them to our model and problem definition, and providing an interpretation of the knowledge that our model can unveil. Of course in both cases we are in front of an approximation. Table \ref{tab:networks} provides the general statistics about the extracted networks for each dataset.

\begin{table}
\centering
\begin{tabular}{|l|rrrrr|}
\hline
Network & $|V|$ & $|E|$ & $|D|$ & $|S|$ & $|D_r|$ \\ 
\hline
DBLP & 38,942 & 100,983 & 16 & 50 & 1,187 \\
Enron & 5,913 & 49,058 & 7 & 8 & 0 \\
\hline
\end{tabular}
\caption{The statistics of the extracted networks.\vspace{-0.5cm}}\label{tab:networks}
\end{table}

DBLP\footnote{\url{http://www.dblp.org/db/}} is an online bibliography containing information about scientific publications in the field of computer science. Using the data from this dataset, our problem definition may be adapted as follows: we want to evaluate the actual knowledge possessed by scientific authors in different topics and sub-topics, focusing on different branches of their disciplines. Being our aim to rank authors, they should be the nodes of our network. The link is the co-authorship relation: two authors are connected if they have written together a paper. The dimension of the connection should represent the ``quality'' of their relation, in this case the venue where the publication appeared (we chose 16 top-tier conferences in computer science, including VLDB, SIGKDD, CIKM, ACL, SIGGRAPH and others). The set of skills should describe the expertise of the author, therefore we chose to represent them as the keywords used in their publications. We eliminated stopwords, we applied a stemming algorithm \cite{Porter80} on the remaining words and then we selected the 50 most commonly used keywords in a paper title. The number of times author $u$ used the keyword $s$ is used to evaluate how much the author considers himself an expert over $s$, i.e. it is used as its $w$ value for $s$.

Enron dataset\footnote{\url{http://www.cs.cmu.edu/~enron/}} is a collection of publicly available emails exchanged by the employees of the energy company, distributed after the well known bankruptcy case. We are interested in ranking the employees, that are the nodes of our network. With this network we are able to unveil who are the real knowledge gateways in an organization, by looking at the internal communication even in the absence of more structured social information (see Section \ref{sec:rankings} for the results). Therefore, to apply our algorithm is not necessary for an organization to actually create a social media platform for their employees (or to download information from other social media). We took only the email addresses ending with ``@enron.com''. We connected two employees if they wrote to each other at least once. Then we used as dimensions the day of the week when the communication took place (ending up with seven dimensions from Monday to Sunday). For the set of skills, we considered the 8 most used keywords in the subject field of the emails (again eliminating stopwords and stemming the remaining words and directly evaluating the relation between an employee and the keyword by the number of times she used the word in an email subject).

\section{The UBIK Algorithm}\label{sec:UBIK}
In this section we discuss the implementation details of our algorithm. We called it UBIK (``you (U) know Because I Know''). UBIK requires the following input: a network $G=(V,E,D,S)$ with the characteristics presented in Section \ref{sec:model}; a range parameter $\alpha$ regulating how much information is lost after each degree of separation; and a set specifying, for each $d \in D$, what is the relevance $r(d)$ of $d$ (defined by the analyst accordingly to the ranking aims).

\begin{algorithm}[t]
\small
\begin{algorithmic}[1]
\REQUIRE{$G=(V,E,D,S); \alpha  \in \left[0\ldots\infty\right]$; $r(D)$}
\ENSURE{Node set $V$ with updated skill values.}
\smallskip
\STATE $\ell \leftarrow 1$
\WHILE{$\ell < \delta$}
\FORALL{$u\ \in\ V$}
\FORALL{$v\ \in\ N(u)$}
\FORALL{$s\ \in\ S$}
\STATE $w'_{u, s} \leftarrow w_{u, s} + \sum_{d \in D} \frac{(f(v, s) \times r(d))^{\frac{1}{\ell \alpha}}}{|N(u)| + |N(v)|}$
\ENDFOR
\ENDFOR 
\ENDFOR
\STATE $UPDATE(V, u'(s))$
\STATE $\ell \leftarrow \ell + 1$
\ENDWHILE
\STATE $NORMALIZE(V)$
\STATE {\bf return} $V$
\end{algorithmic}
\caption{The pseudo-code of UBIK.}
\label{alg:pseudocode}
\end{algorithm}

Aim of UBIK is to update $\forall s \in S$ and $\forall u \in V$ the value $w$, i.e. how much node $u$ possesses of skill $s$. The pseudocode of UBIK is Algorithm \ref{alg:pseudocode}. UBIK cycles for each node, using the following master equation:

$$ f(u, s) = \sum_{d \in D} \sum_{v \in N(u, d)} \dfrac{(f(v, s) \times r(d))^{\frac{1}{\ell \alpha}}}{|N(u)| + |N(v)|} $$

where $N(u, d)$ is a function returning all the neighbors of $u$ that are reachable through dimension $d$ (if a dimension is not specified, it returns the entire neighborhood). Notice that the contribution of each $d$ is different, corrected with the value $r(d)$, i.e. the relevance of dimension $d$. Also note that, at the first iteration, $f(v, s)$ (i.e. how much node $v$ possesses of skills $s$) is equal to just the weight $w_{v, s}$, but at the second iteration it will be updated with the master equation.

One important caveat must be discussed about the $\ell$ parameter. In our model (Section \ref{sec:model}) we said that $\ell$ represent the degrees of separation of the nodes $u$ and $v$, exchanging their skill values. Therefore, the exact implementation of our model would require to scan for each $u$ each node of the network, calculate the shortest path between the two, and then update the contribution accordingly to the $\ell$ value. However, this implementation is inefficient, as it is an equivalent of finding all the shortest paths in the network (that is a cubic problem in terms of the number of nodes, or $|V|^3$ \cite{spalgorithms}) and then apply our calculation. Instead, Algorithm \ref{alg:pseudocode} provides an approximation of the result. The approximation reduces the main loop time complexity as linear in terms of number of edges, usually approximated as $|V| \log |V|$.

We set $\ell = 1$ and we apply the master function to every node and every skill. Then, we increase $\ell$ by 1 and we apply the master function again, using not the original skill values of nodes, but the ones updated at the first iteration. In this way, all the neighbors of $u$ are passing to $u$ also the skills that they have inherited from their neighbors. We avoid to pass back to $u$ the skills that $u$ itself passed to its neighbors at the previous iteration. At the $n$-th iteration, the neighbors of $u$ pass to $u$ the skill values obtained by the nodes $n-1$ degrees away.

The stop criterion is dependent on the $\ell$ value. On average, nodes that are beyond three or four degrees of separation cannot influence significantly the skills accessible from one node. Therefore, in Algorithm \ref{alg:pseudocode}, at step 2 we stop if $\ell \geq \delta$, with $3 \leq \delta \leq 6$, dependent on the application.

When we calculate the new skill values at step 6, we store the result in a temporal variable for each node. Then, we apply the $UPDATE$ function at step 10 to update the value of skill $s$ for node $u$ in the node set $V$. Each element of the master equation is either fixed ($\alpha$, $D$, $r(d)$, $N(u)$ and $N(v)$ are always the same) or it only depends on the previous iteration ($\ell$, $u(s)$ and $v(s)$). By forcing this condition, UBIK becomes order-independent: the computed value of each $u(s)$ at a particular iteration is always the same, regardless if $u$ was considered as the first node of the iteration or as the last.

The $NORMALIZE$ function at step 13 scales for each skill the values obtained for each node in the $[0,1]$ interval. Moreover, it combines all the skill values in a general network ranking. The general value of node $u$ is evaluated by simply extending the $u(s)$ function in the following way: $u(*) = \sum_{s \in S} u(s),$ where $*$ symbolizes the sum of all $s_1, s_2, \ldots, s_n \in S$. This function is run at the end of the main UBIK loop and does not add any complexity to it.

\subsection{Time Complexity}
Algorithm \ref{alg:pseudocode} has five nested loops. From the inner to the outer, they cycle over: the set of dimensions (the sum at step 6), the set of skills (step 5), the neighbors of a node (step 4), the nodes of the network (step 3) and until the $\ell < \delta$. Since cycling over the nodes and their neighbors (steps 3-4) is equivalent to cycle twice over the edges, the complexity of those two loops is $\mathcal{O}(|E|)$. Steps 5-6 have complexity of $\mathcal{O}(|S|)$, while the outer loop generally terminates after very few iterations: in real world networks, usually $\delta = \log |V|$. The final estimate for the time complexity is then $\mathcal{O}(\log|V|\times|E|\times|S|)$. We also report that usually for real world networks, the number of skills and dimensions (both in the order of $10^1$ or $10^2$) is usually much lower than the number of edges (usually ranging from $10^5$ to $10^8$ and more), making the average case complexity in the order of $\Theta(\log |V| \times |E|)$.

\section{Experiments}\label{sec:experiments}
\begin{figure}
\centering
\subfloat[First series.]{\includegraphics[width=0.475\columnwidth]{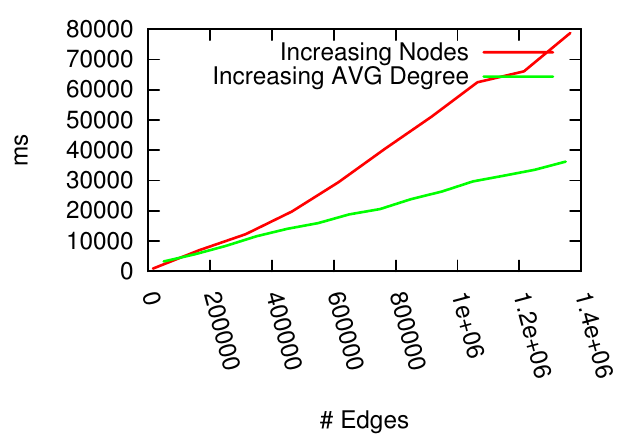}}
\subfloat[Second series.]{\includegraphics[width=0.475\columnwidth]{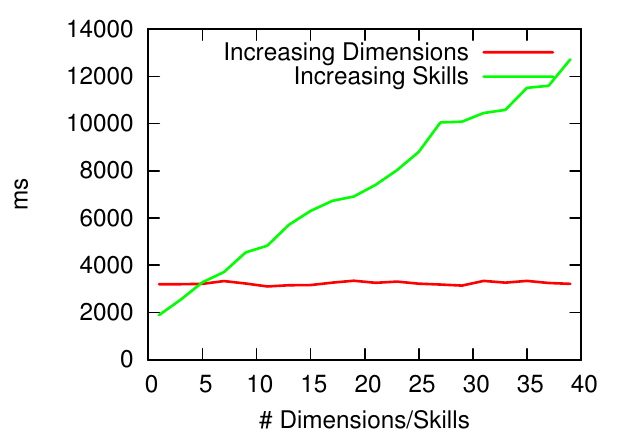}}
\caption{Running times in milliseconds for different random networks with given number of edges, dimensions or skills.\vspace{-0.75cm}}
\label{fig:runtimes}
\end{figure}

We tested our Java implementation of UBIK\footnote{Freely available with our test datasets at \url{http://www.michelecoscia.com/?page_id=480}}, on a Dual Core Intel i7 64 bits @ 2.8 GHz, 8 GB of RAM and a kernel Linux 3.2.0-23-generic, using as virtual machine the Java OpenJDK version 1.6.0\_24. Our implementation took on average 22 seconds on DBLP and less than 2 seconds on Enron. Our networks are small in scale, thus we created some benchmark networks to show how UBIK scales in terms of number of nodes, average degree, dimensions and skills. The results are depicted in Figures \ref{fig:runtimes}a and \ref{fig:runtimes}b.

First, we fixed the number of dimensions and of skills at 5. Then for the ``Increasing Nodes'' series we fixed the average degree at 3 and we increased the number of nodes; while for the ``Increased AVG Degree'' series we fixed the number of nodes at 50k and we increased the average degree of the nodes. Both techniques increase the number of edges: UBIK is able to scale linearly in this dimension. UBIK is able to analyze a network with 455k nodes and 1.3M total edges over all dimensions in less than a minute and a half, or with 50k nodes and the same number of edges in less than 40 seconds. The difference in the linear slope between the two is given by the increasing of both nodes and edges. We can conclude that UBIK is scalable and applicable to large scale networks, as it is linear on the number of edges. In Figure \ref{fig:runtimes}b we fixed the number of nodes at 25,000 and the average degree at 3. Then for the ``Increasing Dimensions'' series we increased the number of dimensions from 1 to 40, while for the series ``Increasing Skills'' we increased the number of skills from 1 to 40. Our implementation, paying a preprocessing phase, is independent from the number of dimensions, the runtime increases linearly with the number of skills\footnote{To assure repeatability, also the random network generator is provided at the same page of the algorithm and the networks}.

We now proceed to evaluate the results of UBIK, comparing its results with some of the state-of-the-art node rank approaches (in Section \ref{sec:comparison}) and presenting some knowledge extraction from real world networks (in Section \ref{sec:rankings}).

\subsection{Comparison with other methods}\label{sec:comparison}
We compare the rankings provided by UBIK with some state-of-the-art algorithms. The algorithms used for comparison are the Personalized PageRank \cite{ppr} and TOPHITS, a tensor eigenvector-based approach to ranking. Personalized PageRank is implemented in the R statistical software, TOPHITS is part of the Tensor Toolbox for MatLab \cite{tophits}, freely available for download\footnote{\url{http://www.sandia.gov/~tgkolda/TensorToolbox/index-2.5.html}}. For our comparison, we used the DBLP network.

We used UBIK without giving to any dimension any particular value of $r(d)$ and we took the global ranking of the nodes without selecting any particular skill. In this way, the comparison with PageRank and TOPHITS is fair, because we are evaluating the general rank of our nodes without using anything else than the network structure, that both the Personalized PageRank and TOPHITS can handle. Also, we set $\alpha = 2$ and $\delta = 6$.

The task of confronting different ranking methods is not easy, as it is not explicit why a ranked list is better than a different one on absolute terms. However, there are several properties that we would like to have in the results of a ranking algorithm. We evaluate the results of the algorithm based on quantitative tests on the following properties:

\begin{enumerate}
\item Ranking results should not be trivial: if the results are highly correlated with a trivial ranking method, then the algorithm is not telling us something interesting.
\item Ranking results should not be trivially boosted: if there is a simple mechanism to increase one's rank, the flaw of the algorithm makes its results less important.
\item Given a gold standard calculated in an independent way, a ranking algorithm is good if it is quantitatively similar, to some extend, to the gold standard.
\end{enumerate}

Let us start from the first element of the list. One of the most trivial criterion for ranking nodes in a network is to check their degree: the more edges are connected to a node, the more important it is. Of course, this ranking method is not optimal, as it takes only an artificial creation of many edges centered on a node to obtain the maximum rank (as it happens in the World Wide Web). Therefore, the most similar to the degree ranking are the results of the algorithm, the less interesting they are. This is only the first test to be satisfied, but it is necessary to satisfy also the other two. For example, a ranking method that uses the inverse of the degree to rank node will pass this test, as it anti-correlates with the trivial degree ranking method, but it will not satisfy the other two conditions.

\begin{figure}
\centering
\subfloat[UBIK.]{\includegraphics[width=0.32\columnwidth]{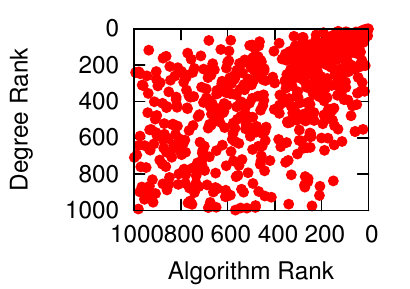}}
\subfloat[PageRank.]{\includegraphics[width=0.32\columnwidth]{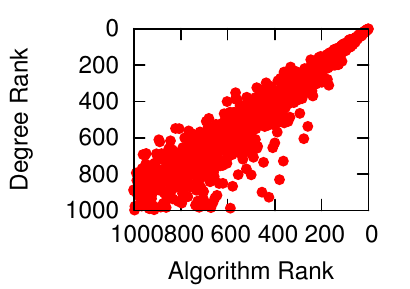}}
\subfloat[TOPHITS.]{\includegraphics[width=0.32\columnwidth]{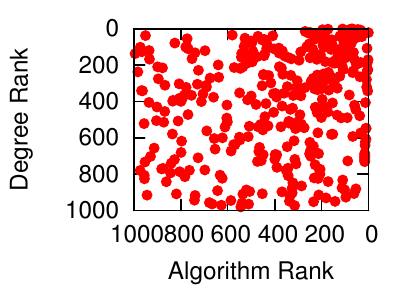}}
\caption{The q-q plots of various ranking algorithms against the ranking obtained ordering the nodes by degree.\vspace{-0.5cm}}
\label{fig:q-q}
\end{figure}

\begin{table}
\centering
\begin{tabular}{|r|lll|}
\hline
R & Degree & PageRank & UBIK\\ 
\hline
1&	Jiawei Han&	Jiawei Han&	Philip S. Yu\\
2&	Philip S. Yu&	Philip S. Yu&	Jiawei Han\\
3&	Christos Faloutsos&	Christos Faloutsos&	Qiang Yang\\
4&	Qiang Yang&	Qiang Yang&	Hans-Peter Kriegel\\
5&	Divesh Srivastava&	Divesh Srivastava&	Gerhard Weikum\\
6&	Zheng Chen&	Jian Pei&	Divesh Srivastava\\
7&	Jian Pei	&Zheng Chen&	Zheng Chen\\
8&	Raghu Ramakrishnan&	Hector Garcia-Molina	&Elke A. Rundensteiner\\
9&	Beng Chin Ooi&	Beng Chin Ooi&	C. Lee Giles\\
10&	Hector Garcia-Molina&	Gerhard Weikum&	Christos Faloutsos\\
11&	Haixun Wang&	Raghu Ramakrishnan&	Wei-Ying Ma\\
12&	Wei-Ying Ma&	Haixun Wang&	Yong Yu\\
13&	Gerhard Weikum&	Wei-Ying Ma&	Tao Li\\
14&	Michael J. Carey&	Michael Stonebraker&	Ming-Syan Chen\\
15&	Jeffrey Xu Yu&	Rakesh Agrawal&	Jian Pei\\
\hline
\end{tabular}
\caption{The top 15 researchers according to different ranking criteria.}\label{tab:q-q-table}
\end{table}

Figure \ref{fig:q-q} depicts the q-q plots of UBIK, PageRank and TOPHITS against the degree ranking for the 1,000 nodes with highest degree. Each point $x(i,j)$ of a q-q plot corresponds to some node $x$. The coordinates of the point $(i,j)$ mean that the node is ranked at the $i$-th position by the first algorithm (x-axis) and at the $j$-th position by the second algorithm (y-axis). In Figure \ref{fig:q-q}, the y axis is the degree rank, while on the x axis we have UBIK (Figure \ref{fig:q-q}a), PageRank (Figure \ref{fig:q-q}b) and TOPHITS (Figure \ref{fig:q-q}c). The interpretation of the picture is clear: especially for the 300 highest ranked nodes, having a high degree implies having a high PageRank, while this consideration does not hold for UBIK and TOPHITS results. We also report the top 15 researchers in Table \ref{tab:q-q-table} for UBIK and PageRank (TOPHITS omitted due to lack of space, but the interesting confront of the table is with the PageRank algorithm, as TOPHITS does not show the rank-degree correlation). Again, we can easily see the correlation between the Degree and the PageRank column. The rank-degree correlation for PageRank is not our finding, as it has already been studied in literature \cite{gourab}. In practice, the logic of the degree centrality is ``The more collaborators a researcher has, the more important he is''. Both PageRank and UBIK modify this philosohpy in ``The more important collaborators a researcher has, the more important he is''. However the ``important'' in PageRank is still a quantitative measure on the number of collaborations, while UBIK uses more qualitative information. For the first criterion, we conclude that UBIK rankings are less trivial than the ones returned by PageRank.

\begin{table}
\centering
\begin{tabular}{|l|rrrrr|}
\hline
Algorithm & (1) & (2) & (3) & (4) & (5)\\
\hline
UBIK & 0.0204 & 1\% & 1.2\% &  4\%& 5.5\%\\
PageRank & 0.0166 & 0\% & 0.8\% & 2\% & 5.3\%\\
TOPHITS & 0.0857 & 39\% & 33.6\% & 34.8\% & 37.5\%\\
\hline
\end{tabular}
\caption{The share of high clustering nodes in the top rankings per algorithm.\vspace{-0.5cm}}\label{tab:k-plot}
\end{table}

So far we have shown the main defect of PageRank, i.e. it provides a trivial ranking. What is the main defect of TOPHITS? In literature, it is studied as the Tightly-Knit Community (TKC) effect: being the TOPHITS rank self-enforced through eigenvector calculation, if we highly clustered nodes in the network, it may happen that all members of this group are ranked high by TOPHITS, even though the nodes are not particularly central. This is the second point we want to prove: UBIK rankings are not prone to these easily applicable ranking boost strategies.

Table \ref{tab:k-plot} reports the share of high clustering nodes for the top k ranked nodes, accordingly to UBIK, PageRank and TOPHITS. The column (2) reports the percentage of nodes in the top 100 ranked by the algorithm that have a local clustering $k > .1$, the column (3) reports the same statistic for the top 250 nodes, and so on. The local clustering $k(i)$ of a node $i$ is defined as:

$$k(i) = \frac{2|\{(u, v) \mid u, v \in N(i) \wedge (u, v) \in E\}|}{|N(i)| \times (|N(i)| - 1)}$$

(note that we calculate the monodimensional clustering, without specifying a $d$ for $N(i)$). We can see that both UBIK and PageRank tend not to return high ranks for nodes with a high local clustering value. On the other hand, in the TOPHITS ranks 39 nodes out of the most important 100 have high clustering values. Table \ref{tab:k-plot} also reports the average local clustering value for the top ranked 100 nodes in column (1) and again this value is $> 4 \times$ higher for TOPHITS. We conclude that both UBIK and PageRank are not affected by the TKC, and that UBIK is the only example that is not dependent both on degree and local clustering.

\begin{table}
\centering
\begin{tabular}{|l|@{\hspace{1mm}}r@{\hspace{1mm}}r@{\hspace{1mm}}r@{\hspace{1mm}}r@{\hspace{1mm}}r@{\hspace{1mm}}r@{\hspace{1mm}}r@{\hspace{1mm}}r@{\hspace{1mm}}r@{\hspace{1mm}}r@{\hspace{1mm}}|}
\hline
&App.&Search&Stream&Obj.&Analysis&Sys.&User&Model&Network&Context\\
\hline
P. S. Yu&1&3&4&7&4&7&5&6&2&6\\
J. Han&5&1&3&5&3&6&4&3&3&4\\
Q. Yang&11&15&1&13&5&14&8&18&7&10\\
H-P. Kriegel&3&2&5&1&7&12&6&10&16&2\\
G. Weikum&7&9&15&2&1&17&2&11&5&5\\
D. Srivastava&18&5&24&21&11&1&12&2&11&7\\
Z. Chen&17&17&2&20&6&22&1&14&20&17\\
Rundensteiner&2&7&11&14&9&15&14&1&8&11\\
C. Lee Giles&14&16&16&16&10&19&7&23&1&14\\
C. Faloutsos&21&6&34&30&15&28&15&20&30&1\\
\hline
\end{tabular}
\caption{The top 10 researchers according to the general UBIK ranking and their rank for 10 different skills.\vspace{-0.5cm}}\label{tab:multiple-rankings-1}
\end{table}

Let us now address the third important feature that a ranking algorithm should have: the comparison with a quantitative gold standard. In scientific publishing, a useful indicator about the quality and the impact of a researcher is quantified using several different indexes. One of them, the h-index, measures both the amount of publications and citations of an author, and it is logically not related to the co-authorship network. A researcher has an h-index of $h$ if he has at least $h$ publications cited at least $h$ times. We use the h-index as our ground truth.

For this comparison we could use a q-q plot, but we need a more quantitative and objective measure than simply looking at the plot. Therefore, we follow \cite{sidiropoulos} and we use a function computing the distance of the points in a q-q plot from the line $y = x$ that represent identical rankings. The distance of point $(i, j)$ from the line $y = x$ is equal to $\frac{|i - j|}{\sqrt{2}}$. Thus, the distance measure $D$ of two rankings $r_1$ and $r_2$ is:

$$ D(r_1, r_2) = \dfrac{1}{|V|} \sum_{\forall v \in V} \frac{|r_1(v) - r_2(v)|}{|V|} $$

where $r_x(v)$ is the rank of $v$ according to the ranking $r_x$.

We calculate this function for UBIK, PageRank and TOPHITS against h-index ranking. We obtained the h-index values from an updated webpage who is collecting data from Google Scholar\footnote{\url{http://www.cs.ucla.edu/~palsberg/h-number.html}}. From that list, we removed the authors that have not published a single paper in the set of conferences with which we have built our multidimensional network, because they are not part of the network structure at all. UBIK's ranking is closer to the ground truth, computed independently from the network structure, provided by the h-index ranking, with a score of 22.43. Therefore, not only UBIK is not affected by the biases of PageRank and TOPHITS, but it also yields results closer to an independent ground truth, as they scored 23.50 and 37.54 respectively. We can now take a look to the actual multiskill and multidimensional rankings provided by the algorithm, as the comparison section is over and we can use the features, not handled by PageRank nor TOPHITS.

\subsection{Rankings}\label{sec:rankings}
We now report some rankings extracted with UBIK. We already saw in Table \ref{tab:q-q-table} the top 15 researcher setting no particular dimension weight (i.e. $r(d)$ is equal for all $d \in D$). However, now we want to take advantage of the fact that UBIK is able to return different rankings for each skill. In Table \ref{tab:multiple-rankings-1} we report the list of researchers who can master some skills, and their ranking for the other skills. As we can see, no researcher dominates over all the skills, and the different rankings can enlighten us about different leaders in different sectors. We remind that we took only authors of a very specific set of conferences, thus a possible specialist in one or more reported skills may not be part of the rankings because she never published in one of the selected conferences. Also, the skill name is the substantive of the stemmed form, thus it includes all the possible declination of the term (e.g. ``Stream'', ``Streaming'', ``Streamed'', and so on).

UBIK is able to customize the ranking even further, in an additional degree of freedom. Instead of looking at some skills taken separately, we can populate our set of dimension relevance functions with different importance $r(d)$ values for different dimensions. A proper definition of the dimension importance set results in a very specific ranking analysis. To show this feature, we decide to create two different definition classes for the Enron network. We recall that in the Enron network each dimension is the day in the week when the email was sent. In the first variant, we populate our set of rules with a $10\times$ multiplier for the dimensions of Saturday and Sunday; in the second variant we apply the same $10\times$ multiplier, but this time to each of the weekday, and nothing to the weekend days. The choice of the $10\times$ multiplier is \textit{ad hoc}, to represent a query that focuses on the relations established mainly during weekdays (first case) or during weekends (second case).

\begin{table}
\centering
\begin{tabular}{|l|rr|}
\hline
Employee&Weekday Rank&Weekend Rank\\
\hline
Victor Lamadrid&1&34\\
Jeff Dasovich&2&4\\
Lisa Jacobson&3&39\\
Michael Kass&4&2\\
Joannie Williamson&5&37\\
Bob Ambrocik&6&68\\
Chris Germany&7&8\\
Kay Chapman&8&27\\
Tana Jones&9&23\\
Drew Fossum&10&18\\
Forrest Lane&238&1\\
Jennifer Blevins&1760&3\\
Shubh Shrivastava&857&5\\
Kay Mann&14&6\\
Scott Neal&50&7\\
Vince Kaminski&18&9\\
Rosalee Fleming&21&10\\
\hline
\end{tabular}
\caption{The top 10 employees according to the UBIK for the weekday and the weekend variants of the ranking.\vspace{-0.5cm}}\label{tab:multiple-rankings-2}
\end{table}

Table \ref{tab:multiple-rankings-2} reports the top 10 employees according to both criteria (please note that some employees are part of both top 10 and they are not repeated). As we can see, the two rankings are quite distinct. There are three employees very important in both criteria (Jeff Dasovich, Michael Kass and Chris Germany). We also observe one expected phenomenon: the important employees during the weekdays are also somewhat important during the weekends, while the vice versa is not true. It is expected that important employees receive emails during the entire week, while the communications during the weekend may follow a different logic and promote unexpectedly low ranked employees (maybe because they perform a weekend shift or due to particular emergencies outside office hours).

The most notorious elements of the top management of Enron are not present in either rankings. Kenneth Lay is ranked 617th in weekdays and 224th in weekends, while Jeffrey Skilling is ranked 725th in weekdays and 458th in weekends. Joannie Williamson, who worked as a secretary for both of them\footnote{\scriptsize \url{http://money.cnn.com/2006/04/03/news/newsmakers/enron\_defense/index.htm}}, is instead present in Table \ref{tab:multiple-rankings-2}, and highly ranked. This is an expected result of UBIK, able to unveil who is a knowledge gateway in an organization.

\begin{table}
\centering
\scriptsize
\begin{tabular}{|l|rrrrrrr|}
\hline
Author&ACL&CIKM&ICDE&KDD&GRAPH&VLDB&WWW\\
\hline
D. Marcu&1&3893&3511&2606&2309&1608&3327\\
Boughanem&2768&1&3892&2966&2643&1946&3658\\
T. Ichikawa&3070&4549&1&3266&2982&2260&3994\\
Nakhaeizadeh&2606&4036&3709&1&2484&1783&3507\\
D. Salesin&2101&3538&3149&2304&1&1300&2980\\
P. Dubey&3093&4570&4198&3274&2988&1&3999\\
J. Nieh&2976&4453&4134&3185&2865&2158&1\\
\hline
P. S. Yu&574&32&27&10&860&10&684\\
J. Han&689&106&60&12&999&29&708\\
Q. Yang&933&634&967&384&1258&232&930\\
H-P. Kriegel&1024&517&66&330&1304&146&1476\\
G. Weikum&1100&472&91&888&1348&51&1355\\
\hline
\end{tabular}
\caption{The top authors for some conferences (taken singularly) comparative rankings.\vspace{-0.5cm}}\label{tab:multiple-rankings-3}
\end{table}

The same multidimensional ranking can be done for the DBLP network. In this case, we are able to spot the collaboration hubs in several different conferences. We applied the $10\times$ multiplier to a collection of conferences. The results for some of our conferences are provided in Table \ref{tab:multiple-rankings-3}, where for each conference we record the top ranked author and then we report also his ranking for the other conferences. We can see that UBIK is able to identify specialists who are highly ranked only in one conference. On the other hand, as expected, the top authors in the general ranking score average high rank in all conference, but they are rarely in the top 10 of a specific conference, as their impact is more broad and it spreads over many different venues (the bottom rows of Table \ref{tab:multiple-rankings-3} records the ranking for each single conference for the top 5 general authors in the network taken without any dimension multiplier).

\section{Conclusion}\label{sec:conclusion}
In this paper we addressed a problem of human resources: the ranking of employees according to their skills. We did so following an intuition about the intellectual value of a person: her evaluation should not be based only on her set of skills, but also on her friends' set of skills. We created an algorithm, UBIK, to tackle this problem: UBIK is able to rank nodes in a multidimensional network, with weighted labels referring to their set of skills. We applied UBIK to two real world networks, confronting its output with popular ranking algorithms, showing that our results are less trivial and more significant. Our contributions are: the creation of a feasible tool to address a problem in organizations in the real world; the address of the multiple ranking over multidimensional networks problem, not tackled in literature; and increased ranking performances over the current most used methodologies.

Our work opens the way to a number of future works. First, following \cite{multirank}, we can extend UBIK to rank also the relations. UBIK can be applied to several different datasets with different semantics and properties, from Linkedin to evaluate people's expertise; to networks of international organizations, to detect the most important organizations given a set of topics. Lastly, we can create a more efficient implementation, confront with other algorithms and evaluation measures.

\textbf{Acknowledgements}. This work has been partially supported by the European Commission under the FET-Open Project n. FP7-ICT-270833, DATA SIM.

\balance

\bibliographystyle{IEEEtranS}

\bibliography{asonam2013}

\end{document}